\RequirePackage{fix-cm}
\documentclass[smallcondensed,natbib]{svjour3}     
\smartqed  
\usepackage{graphicx}
\usepackage{subfigure}
\usepackage[utf8]{inputenc}
\usepackage{booktabs}
\usepackage{rotating}
\usepackage{hyperref}
\usepackage{amssymb}
\usepackage{hyphenat}
\usepackage{textcomp}
\usepackage[resetlabels]{multibib}
\newcites{S}{Included Secondary Studies}

\journalname{Software Quality Journal}
\begin{document}

\title{Quality in Model-Driven Engineering}
\subtitle{A tertiary study}

\titlerunning{Quality in Model-Driven Engineering: a tertiary study}        

\author{Miguel Goulão   \and
        Vasco Amaral    \and
        Marjan Mernik
}

\authorrunning{Miguel Goulão, Vasco Amaral and Marjan Mernik} 

\institute{Miguel Goulão \at
              NOVA LINCS, Departamento de Informática\\
              Universidade Nova de Lisboa\\
              Tel.: +351-21-2948536\\
              Fax: +351-21-2948541\\
              \email{mgoul@fct.unl.pt}
           \and
           Vasco Amaral \at
              NOVA LINCS, Departamento de Informática\\
              Universidade Nova de Lisboa\\
              Tel.: +351 212948536\\
              Fax: +351 212948541\\
              \email{vma@fct.unl.pt}
           \and
           Marjan Mernik \at
            Faculty of Electrical Engineering and Computer Science\\
            University of Maribor\\
            Tel.: +386 22207455\\
            Fax: +386 22207272\\
            \email{marjan.mernik@um.si}
        }

\date{Received: date / Accepted: date}

\maketitle

\begin{abstract}
\textit{Context:} Model-Driven Engineering (MDE) is believed to have a significant impact in software quality. However, researchers and practitioners may have a hard time locating consolidated evidence on this impact, as the available information is scattered in several different publications.\\
\textit{Objective:} Our goal is to aggregate consolidated findings on quality in MDE, facilitating the work of researchers and practitioners in learning about the coverage and main findings of existing work as well as identifying relatively unexplored niches of research that need further attention.\\
\textit{Method:} We performed a tertiary study on quality in MDE, in order to gain a better understanding of its most prominent findings and existing challenges, as reported in the literature.\\
\textit{Results:} We identified 22 systematic literature reviews and mapping studies and the most relevant quality attributes addressed by each of those studies, in the context of MDE. Maintainability is clearly the most often studied and reported quality attribute impacted by MDE. 80 out of 83 research questions in the selected secondary studies have a structure that is more often associated with mapping existing research than with answering more concrete research questions (e.g. comparing two alternative MDE approaches with respect to their impact on a specific quality attribute). We briefly outline the main contributions of each of the selected literature reviews.\\
\textit{Conclusions:} In the collected studies, we observed a broad coverage of software product quality, although frequently accompanied by notes on how much more empirical research is needed to further validate existing claims. Relatively little attention seems to be devoted to the impact of MDE on the Quality in Use of products developed using MDE.\\

\keywords{Quality \and Model-Driven Engineering \and Tertiary Study}
\end{abstract}


%
%

\section{Introduction}
\label{sec:Introduction}

\textbf{M}odel-\textbf{D}riven \textbf{E}ngineering (MDE) \citep{Schmidt2006Computer,Silva2015COMLAN} is an increasingly popular software development methodology that makes use of rigorous abstractions, known as models (that allow predictions or inferences to be made), to design and implement a given system. Models are used to fight the complexity of specifying systems for nowadays target platforms, not tackled by the general purpose programming languages, and express domain concepts effectively. Therefore, together with model transformation approaches (i.e. model-to-model and model-to-code), that promote automatic or semiautomatic generation of code, models are no longer just means of documentation but they are a core development artifact of the system. Models are commonly specified in visual modeling languages, e.g. \textbf{U}nified \textbf{M}odeling \textbf{L}anguage (UML) \citep{OMG2015UML25}, but can also be specified in dedicate textual languages, like Domain Specific Languages, as long as they are able to abstract concepts from the domain of the problem instead of the domain of the solution (e.g. RubyTL \citep{cuadrado2006rubytl}).

MDE is often claimed to bring several benefits to software projects that, ultimately, allow dealing with the increasing development complexity of the software systems, leading to cost reductions in the development and evolution processes, while also increasing the quality of the target developed software. The apparent relatively low level of adoption of MDE in industry does not seem to be in line with these benefits. A growing number of studies (mostly surveys with practitioners) is shedding light to the actual usage of MDE, as well as some of the challenges MDE is still facing, hampering a wider adoption by practitioners. In this paper we are interested to observe the status of both perspectives regarding quality of the product and process. 

Some reports point to a relatively low adoption level of modeling (defined as the activity of designing system models in a given language). \citep{Forward2010EEMDD} notes that models are mostly created for documentation and up-front design, but not so much for code generation partly due to the predominance of code-centric environments, rather than modeling-centric ones. More recently, \citep{Hutchinson2014SCP} reported a stronger usage of code generation. Nevertheless, the prevalence of code-centric projects is visible, for example, in open-source projects \citep{Badreddin2013OSSQV}. Although the UML is often claimed to be a \textit{de facto} standard in industry, a recent survey reported that only 15 out of 50 practitioners use it \citep{Petre2013ICSE}. Another survey \citep{FernandezSaez2015MODELS} focused on how UML documentation is used to support maintenance tasks in industry and found a wider adoption level (43\% of the respondents use it, and an additional 16\% use an alternative language for that purpose, with a set of 178 practitioners from 12 different countries). In some countries, like Italy, MDE is becoming much more popular, with 13\% of companies working with modeling always, and an additional 53\% doing it sometimes, in a survey conducted with 155 practitioners \citep{Torchiano2013JSS}. 

There may be several reasons contributing to these challenges when it comes to MDE adoption. Thanks to the technological advances and scientific investment in MDE in the last decade, we are now reaching a level of maturity that allows us to observe the current status and understand its impact in systems development and understand what can be hampering a wider adoption. \citep{Whittle2015SoSyM} reports a summary of interviews with the practitioners' perceptions about the challenges of the current state of practice with MDE tools. Here, several quality factors are recurrently mentioned, namely: the quality of the generated code; the excessive complexity of tools and the abstraction mismatch between the tools offered (including languages) and the users; the impact of the tools in the quality of the product and processes, where productivity and maintainability are the major concerns, e.g. how well tools integrate in the process. \citep{Agner2013JSS} reports a survey on 209 Brazilian software engineers, concluding there is a strong perception of the relevance of MDE in the promotion of quality, although the level of adoption is much lower. The result of this study puts into evidence the quality as a key challenge of MDE. In this paper we will select and consolidate existing work on the impact of MDE in quality. 

Despite the relevance of this perspective we should not ignore other adoption challenges of MDE. For example, \citep{Ameller2015RENEXT} reports on an ongoing international study to assess the way MDE is being used in industry to cope with non-functional requirements, so that the main shortcomings of existing approaches, as perceived by practitioners, can be transformed into significant improvement opportunities.

Others, mention social or organizational challenges \citep{Hutchinson2011ICSE, Hutchinson2014SCP} like resistance to change \citep{Mohagheghi2013EMSE}. \citep{Cuadrado2014SCP} presents lessons learned while introducing MDE in small companies, and notes that most existing studies are performed in medium and large organizations. This is problematic due to the high number of smaller companies which could benefit from adopting MDE, but may find it hard to uncover compelling information supporting that decision.

The challenges of MDE adoption may even have a pedagogic/training nature. Industry representatives have reported the difficulty of hiring well-trained MDE practitioners \citep{Whittle2015SoSyM}. \citep{Badreddin2015HuFaMo} reports on a survey conducted in three universities from the US, Canada and Israel, where students' perceptions on the value of modeling decrease as students progress in their degree training. Additionally, \citep{FernandezSaez2015MODELS} mentions the lower performance with geographically distributed collaborative work. 

All these studies aim, to some extent, to mitigate a perceived problem of lack of consolidated evidence that is compelling enough so that industry can fully understand the strengths, but also the challenges and shortcomings, of MDE adoption. These evidences can be collected from surveys as those reported earlier, or, more generally, from other primary studies on MDE. In this paper, we are particularly interested in the evidence concerning the impact of MDE adoption in the quality of the products built with it (and in the quality of the process of building and evolving those products with MDE). The area is too broad for a single study, but there are already several secondary studies, in the form of literature reviews that partially address it. In this paper we leverage those secondary studies, by conducting a tertiary study to provide an overview of how MDE impacts quality.

The main contributions of this paper are:
\begin{itemize}
\item \textbf{C1:} A mapping of the most representative secondary studies that cover quality in MDE, their origin (to identify key players in this research area), and the quality attributes addressed by each of the secondary studies. This mapping is expected to serve as a starting point for researchers and practitioners interested in quality in MDE to locate relevant consolidated reviews on the area, and the experts responsible for those reviews.
\item \textbf{C2:} An annotated overview of the existing aggregated information on quality in MDE. By outlining the main contributions of the included secondary studies, we provide a single resource from which researchers and practitioners can start exploring further work.
\item \textbf{C3:} A report on the level of consolidation of the aggregated information on quality in MDE. One of the roles that a literature review can play is to aggregate and consolidate information that would be otherwise scattered in several different publications, providing a better overview of the area, while preserving traceability links to the original research sources. In this tertiary review, we are interested in the extent to which this has been achieved in the reported literature reviews.
\end{itemize}

This paper is organized as follows. Section \ref{sec:ResearchMethod} reports the methodology followed to conduct our literature review. Section \ref{sec:Results} reports data collected from each of the selected secondary studies. Section \ref{sec:Discussion} answers our research questions, and reports the limitations of our study. Finally, section \ref{sec:Conclusions} reports our conclusions from this tertiary study.
\section{Research Method}
\label{sec:ResearchMethod}
This paper can be characterized as a work in \textbf{E}vidence-\textbf{B}ased \textbf{S}oftware \textbf{E}ngineering (EBSE) \citep{Kitchenham2004ICSE}. EBSE aims at collecting the best available evidence to address software engineering research questions, both from practitioners and researchers. Typically, this is performed by aggregating existing empirical studies (referred to as primary studies) on a particular topic and performing a literature review on them, in an unbiased way. This is often done through a \textbf{S}ystematic \textbf{L}iterature \textbf{R}eview (SLR). With enough primary studies available, one should aggregate the evidence collected from them to answer the specified research questions (e.g. Is technique \textit{A} more effective than technique \textit{B} under a specific context?), by performing a statistical meta-analysis on those evidences. Unfortunately, it is relatively rare to find good meta-analysis opportunities in software engineering, due to the lack of sufficient primary studies on the phenomenon under scrutiny. Often, we will find more coarse-grained research questions aimed at mapping the existing knowledge on a specific topic rather than aggregating it through meta-analysis. These questions are addressed in \textbf{S}ystematic \textbf{M}apping \textbf{S}tudies (SMSs), which are also a systematic form of literature review. They differ from conventional SLRs in that they are aimed at classifying and performing a thematic analysis of their topic, with more generic research questions, a broader scope, a more inclusive search process, and resulting in a categorization of existing work in several dimensions (see, for example, \citep{Kitchenham2011IST} for a pragmatic comparison between SLRs and SMSs). SMSs can be performed prior to SLRs, to help identifying research questions which are good candidates for data aggregation. Both SLRs and  SMSs are considered secondary studies, as they review primary studies.

This work can be classified as a \textit{tertiary study}. It was conducted following the guidelines on SLRs (and SMSs), detailed in \citep{Kitchenham2007SR}. We can regard this work as a systematic mapping study targeted at secondary studies (\textit{i.e.} SLRs and  SMSs), rather than at primary studies, in order to gather consolidated evidence from those secondary studies, hence being a tertiary study. Our goal, with this decision, was to gather a broader overview on the current state of the art concerning the research and practice of quality in MDE.

\subsection{Research questions}
\label{subsec:ResearchQuestions}
The research questions addressed in this study are:

\begin{itemize}

\item \textbf{RQ1: What is the currently available information concerning quality in MDE, systematically aggregated by means of a SLR, or a  SMS?} Our goal is to characterize the current state of the art in this research domain, providing readers with pointers for looking up additional information concerning each particular quality characteristic and how it relates to MDE.
\begin{itemize}
    \item \textit{RQ1.1: How many primary studies are included in these reviews?}
    \item \textit{RQ1.2: What is the time span covered by these reviews?}
    \item \textit{RQ1.3: Is the list of the included primary studies available?}
    \item \textit{RQ1.4: Is the quality of the primary studies assessed? If so, how is this done?} This assessment is often \textit{ad-hoc}, but can also follow guidelines, such as those in \citep{Kitchenham2007SR}.
    \item \textit{RQ1.5: Who is the secondary study targeted to?} The target audience can be made of \textit{researchers}, \textit{practitioners}, or \textit{both}.
    \item \textit{RQ1.6: What kind of software engineering task is the secondary study aimed at?} These include \textit{decision support}, \textit{knowledge support} and \textit{scoping}.
    \item \textit{RQ1.7: What is the object of study in the included secondary studies?} Typically, this includes \textit{models}, \textit{transformations}, or the \textit{software process} (at least one of these, although for several papers, more than one of these objects of study are addressed).
    \item \textit{RQ1.8: Which quality attributes are addressed in the secondary study?} These include product quality and quality in use attributes, as defined by the ISO/IEC 25010:2011 standards \citep{ISO250102011}.
\end{itemize} 

\item \textbf{RQ2: What is the current status of consolidation of data collected from different literature reviews covering quality in MDE and made available through published secondary studies?} Ideally, one of the potentially key benefits to be taken from literature reviews (in particular, from SLRs) is to be able to aggregate data collected independently in different studies. 

\item \textbf{RQ3: Who are the key players in consolidating knowledge on quality in MDE through literature surveys?} This question is of a demographic nature. Our goal is to characterize the community in this research domain. To answer this question, we break it down into more refined ones, as follows:
\begin{itemize}
    \item \textit{RQ3.1: How many systematic reviews, including systematic literature reviews and mapping studies, addressing the topic of quality in MDE, are available?} This provides a quick overview on the liveliness of the community which is consolidating information on this research topic.
    \item \textit{RQ3.2: What is the origin of these reviews?} This covers organizations and countries, and is  aimed at better understanding the extent to which quality in MDE is becoming a global concern.
    \item \textit{RQ3.3: What is the impact of these reviews, in terms of citations?} This will provide some insight on the ``popularity'' of this research topic.
\end{itemize}

\end{itemize}

\subsection{Search process}
\label{subsec:SearchProcess}
We searched two digital libraries, namely IEEExplore and ACM Digital Library, and the SCOPUS indexing system. All searches were based on title, keywords and abstract. In IEEExplore and ACM digital libraries, full text was also searched. 

The canonical search string, which was adapted for coping with the different syntax rules required by each search engine was as follows:

\begin{verbatim}
("model driven" OR "MDE" OR "MDD") AND
("systematic review" OR "literature review" OR 
 "literature survey" OR "survey" OR 
 "overview of research" OR "mapping study" OR "review")
\end{verbatim}

Our secondary studies selection process started by running the search string in the three repositories. We collected the search results, and merged them into a single spreadsheet, keeping track of the origin of each of the papers. The results yielded 481 candidate papers. These papers were then passed on to the next step of our selection process. 

\subsection{Inclusion and exclusion criteria}
We defined objective inclusion and exclusion criteria, so that the selection process could be conducted in a consistent way. These inclusion and exclusion criteria were applied to the 481 candidate papers.
The inclusion criteria are as follows:

\begin{itemize}
\item \textbf{IC1:} The paper is a peer-reviewed full paper.
\item \textbf{IC2:} The paper includes a literature review where the papers were included using a defined search process -- e.g. a defined search string on an explicit set of digital libraries, or a manual search conducted on the proceedings of a specific venue, within a defined time interval (for example, a manual search on the \textbf{I}nternational \textbf{C}onference on \textbf{S}oftware \textbf{E}ngineering (ICSE) conference series, from 2000 to 2014).
\item \textbf{IC3:} The paper is related to model-driven development, or engineering, and explicitly covers quality in that context.
\item \textbf{IC4:} The paper is written in English.
\end{itemize}

The exclusion criteria provide a complementary view on the inclusion criteria and are as follows:

\begin{itemize}
\item \textbf{EC1:} The paper is not a peer-reviewed full paper. This excludes presentations, extended abstracts, white papers, introductory forewords for conferences or journal special issues, non peer-reviewed book chapters, and books, to name some of the most common examples.
\item \textbf{EC2:} The paper reviews primary studies following an undefined studies selection process.
\item \textbf{EC3:} The paper is out of scope.
\item \textbf{EC4:} The paper is not written in English.
\item \textbf{EC5:} The paper is subsumed by another paper which is already included in our literature review.
\item \textbf{EC6:} The paper is a duplicate of another paper in the sample.
\item \textbf{EC7:} The paper does not include a secondary study.
\end{itemize}

The first exclusion criteria used after the automated search was \textbf{(EC6)}.
As expected, some papers were indexed by more than one repository and, therefore, duplicates. We excluded all duplicates, keeping track of which repositories contained which papers, originally.

Table \ref{tab:PapersFound} summarizes the key statistics on our automatic search, using our search string, followed by the application of \textbf{(EC6)}. We present the individual totals for each of the repositories searched, their sum, the number of removed duplicates, and the total number of candidate papers, after removing duplicates.

\begin{table}[ht]
\centering
\caption{Candidate papers found through automated search}
\label{tab:PapersFound}
\begin{tabular}{@{}cccccc@{}}
\toprule
{\bf ACM} & {\bf IEEExplore} & {\bf SCOPUS} & {\bf Sum} & {\bf Removed duplicates} & {\bf Total}\\ \midrule
127 & 96 & 258 & 481 & -128 & 353\\ \bottomrule
\end{tabular}
\end{table}

We then proceeded to screening the papers titles and abstracts, to detect papers that should remain as candidates for further analysis. Each paper was screened by at least one author. We were very conservative in this first assessment. Whenever we suspected the paper might contain a (possibly systematic) literature review, or mapping study, we kept it for further analysis. This conservative approach was chosen to mitigate the risk of missing important studies. Indeed, it is relatively common for software engineering abstracts to have completeness and clarity shortcomings, as discussed in \citep{Budgen2008EMSE}, which make this screening more error prone, when compared to what can be achieved with structured abstracts, as the one used in this paper. In cases where we were inclined to reject a paper, but not totally confident about this decision, we marked it for a second review by a different author. We only rejected directly papers during title and abstract screening when we were very confident they were not secondary studies, using \textbf{(EC1)}, or when they were obviously out of scope, using \textbf{(EC3)}. This process eliminated a total of 310 papers out of the 353 candidate papers. Of the excluded 310 papers, 42 were not peer review papers (e.g. our search string caught several foreword chapters in conference proceedings), while the remaining papers were out of scope. Out of scope papers include papers on unrelated topics that somehow made it through the automated search and, more commonly, papers that are within our topic, but do not include secondary studies \textbf{(EC7)}, thus being inadequate for inclusion as part of the list of papers scrutinized in a tertiary study. 

This process lead to selecting 43 papers for full text review and, if eligible, for data extraction. All these papers were reviewed by at least one of the authors. 8 of these papers were then excluded for not being a systematic review. Typically, these were informal literature surveys, from where it was impossible to recover the primary study selection process, as expressed by \textbf{(EC2)}. 14 papers were excluded for being out of scope, following \textbf{(EC3)}. These papers would typically have a systematic literature review, or mapping study, but their research questions were not aligned with our own. Finally, 2 papers were excluded because they were preliminary secondary studies later extended in other papers, following \textbf{(EC5)}. We kept the most complete version of these papers for further scrutiny.

In the end of the full text data extraction, we had 19 papers, plus 4 identified through snowball, by checking the references in the selected papers. 1 of the papers identified through snowball was subsumed by a second paper, also identified in the snowball process, so it was discarded. We ended with 22 secondary studies that we will analyze in the remaining of this paper. 

Table \ref{tab:InclusionExclusion} summarizes the included and excluded papers in each phase and the reasons for paper exclusion from our final data set.

\begin{table}[ht]
\centering
\caption{Included and excluded papers in each phase}
\label{tab:InclusionExclusion}
\begin{tabular}{@{}lrrr@{}}
\toprule
{\bf Phase}                                   & \multicolumn{1}{l}{{\bf Included}} & \multicolumn{1}{l}{{\bf Excluded}} & \multicolumn{1}{l}{{\bf Remaining}} \\ \midrule
{\bf Automated search}                        & {\bf 481}                          & {\bf }                             & {\bf }                              \\
{\bf Duplicates removal (EC6)}                & {\bf }                             & {\bf 128}                          & {\bf 353}                           \\
{\bf Title + Abstract screening}              &                                    & {\bf 310}                          & {\bf 43}                            \\
- Not a peer reviewed paper (EC1)             &                                    & (42)                               &                                     \\
- Out of scope (EC3)                          &                                    & (268)                              &                                     \\
{\bf Full-text reading / Data extraction}     &                                    & {\bf 24}                           & {\bf 19}                            \\
- Not a systematic review (EC2)               & \multicolumn{1}{l}{}               & (8)                                & \multicolumn{1}{l}{}                \\
- Out of scope (EC3)                          & \multicolumn{1}{l}{}               & (14)                               & \multicolumn{1}{l}{}                \\
- Subsumed by another paper (EC5)             &                                    & (2)                                &                                     \\
{\bf Snowball search}                         & {\bf 4}                            & {\bf }                             & {\bf 23}                            \\
{\bf Full-text reading / Data extraction}     & {\bf }                             & {\bf 1}                            & {\bf 22}                            \\
- Subsumed by another paper (EC5)             &                                    & (1)                                &                                     \\
{\bf Final set of included secondary studies} & {\bf }                             & {\bf }                             & {\bf 22}                            \\ \bottomrule
\end{tabular}
\end{table}

Concerning the completeness of this paper identification process, we started with informal searches, along with a set of papers we were aware of. After this process, all the previously identified papers had been included, which gives us some confidence on the completeness of this work. Moreover, as we used an additional snowball search step, built on the references lists in the included papers, we expect most of the relevant papers were detected in this aggregated search. All exclusions during the full text reading were double checked by a different author, in order to minimize the risk of missing a relevant paper.

\subsection{Quality assessment}
\label{subsec:MethodQualityAssessment}
The confidence we can place in the results of a systematic review is largely dependent upon the quality of the evidence collected in that review \citep{Dyba2008ESEM,Zhou2015EASE}. The same holds for a tertiary study, such as this one, where appraising the quality of the secondary studies selected for our review is extremely relevant.

The so-called DARE criteria\footnote{The \textbf{DARE} criteria are based on those used by the Center for Reviews and Dissemination of the University of York, for assessing the eligibility of systematic reviews to be included in their \textbf{D}atabase of \textbf{A}bstracts of \textbf{R}eviews of \textbf{E}ffects (DARE). \url{http://www.crd.york.ac.uk/CRDWeb/}} are often used in the context of software engineering -- see, for examples, a relaxed version of DARE in \citep{Kitchenham2010IST} and a complete version in \citep{Cruzes2011IST}). In this paper we use both versions in the appraisal of the quality of the selected papers, keeping the evaluation criteria and scales taken from \citep{Cruzes2011IST,Kitchenham2010IST}, for easier comparison with other tertiary studies adopting these criteria. In the original DARE criteria, used in \citep{Cruzes2011IST}, reviews have to meet at least four of these criteria, and the first three are mandatory. In the more relaxed version of the DARE criteria, used in \citep{Kitchenham2010IST}, only 3 out of 5 criteria are required, with only criteria 1 and 2 as mandatory. We describe those criteria in detail in section \ref{subsec:ResultsQualityEval}, where we then apply them to analyse the quality of the included secondary studies.

\subsection{Data extraction process}
\label{subsec:DataExtractionProcess}
The data extraction process was conducted by the three authors. After selecting the candidate papers, these were split among us, so that one of us would read the paper, extract information from it, and store the collected data in a spreadsheet shared in the cloud by the three authors. Whenever some detail of the data extraction was unclear, we marked it to make sure a second author would double check it. We also kept comments on the shared spreadsheet so that the rationale for decisions and classification of the data collected from the paper was recorded and discussed in our team meetings. In practice, all selected papers were read by at least two of the authors, the first for data extraction and the other(s) for double-checking and helping in situations where the secondary studies information was perceived as less clear. 

\subsection{Extracted data}
\label{subsec:ExtractedData}
In this section we describe the main information gathered from each secondary study:
\begin{itemize}
\item Basic identification information, including authors, their affiliations and countries, source in which the secondary review was published, year of publication and number of citations (taken from Google Scholar).
\item The starting and ending years of publication of the primary studies covered in the secondary study.
\item The number of primary studies included in the secondary study.
\item Information on the availability of the list of included primary studies.
\item Information concerning if and how (i.e. according to which criteria) the quality of the included primary studies was assessed.
\item The target audience (as reported by its' authors) for the secondary study, which may include researchers, practitioners, or both.
\item The secondary study's intended use, as reported by its authors. In particular, we classify secondary studies as being aimed at decision support, knowledge support, or scoping.
\item The objects of study considered in the secondary study, namely models, transformations, and the software process itself.
\item The quality models used as reference in the secondary study, if any.
\item The product quality model attributes addressed in the secondary study, if any.
\item The quality in use model attributes addressed in the secondary study, if any.
\end{itemize}

The last two items this list (quality model and quality in use attributes addressed by the secondary study) were categorized using standard quality models from ISO/IEC. We use the standard ISO/IEC 25010:2011 quality model \citep{ISO250102011}, in particular its model \textit{product quality} and its \textit{quality in use} as references. In both cases, quality concerns both the quality of the products built with MDE and of the tools and methods supporting MDE itself.

\begin{figure}[ht]
 	\centering
    \includegraphics[scale=0.55]{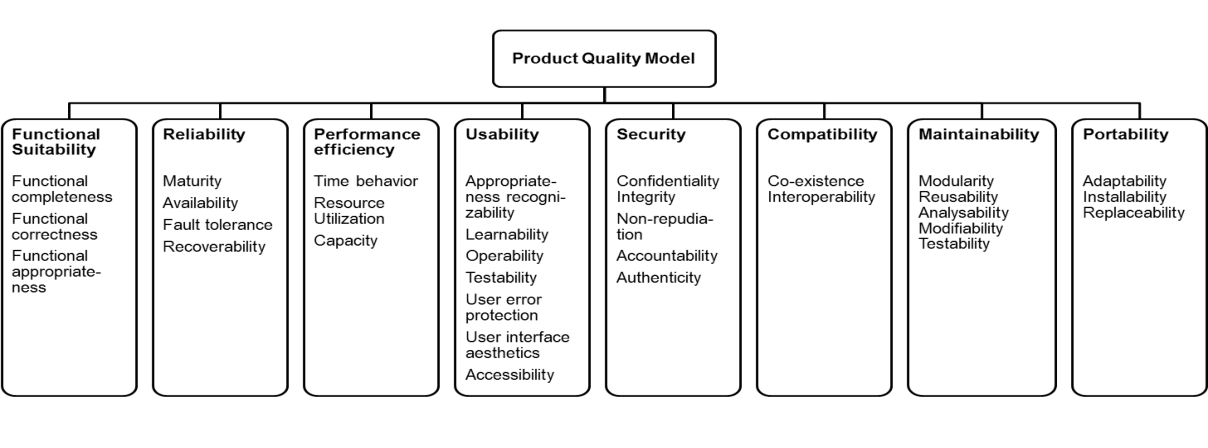}
	\caption{ISO 25010:2011 - Software product quality}
	\label{fig:ProductQuality}
\end{figure}

The product quality model covers both static properties of a software and the dynamic properties of the computer system in which it runs. Using this model allows us to reason about the existing research concerning product quality attributes and how they are regarded in MDE research, according to the selected secondary studies. This quality model, presented in figure \ref{fig:ProductQuality}, includes 8 attributes, which are then further decomposed into sub-attributes. Here we describe the higher level attributes, which we will use to guide our discussion on the coverage of quality attributes in the included secondary studies.

\begin{itemize}
\item \textit{Functional suitability} is the capability of the software product to provide functions which meet stated and implied needs when the software is used under specified conditions.
\item \textit{Reliability} is the capability of the software product to maintain a specified level of performance when used under specified conditions.
\item \textit{Performance efficiency} is the capability of the software product to provide appropriate performance, relative to the amount of resources used, under stated conditions.
\item \textit{Usability} is the capability of the software product to be understood, learned, used and attractive to the user, when used under specified conditions.
\item \textit{Security} is the capability of the software product to protect information and data and controls the access level of people, products, or systems.
\item \textit{Compatibility} is the capability of a product, system, or component, to exchange information with other products, systems and components.
\item \textit{Maintainability} is the capability of the software product to be modified.  Modifications may include corrections, improvements or adaptation of the software to changes in environment, and in requirements and functional specifications. 
\item \textit{Portability} is the capability of the software product to be transferred from one environment to another.
\end{itemize}

\begin{figure}[ht]
 	\centering
	\includegraphics[scale=0.28]{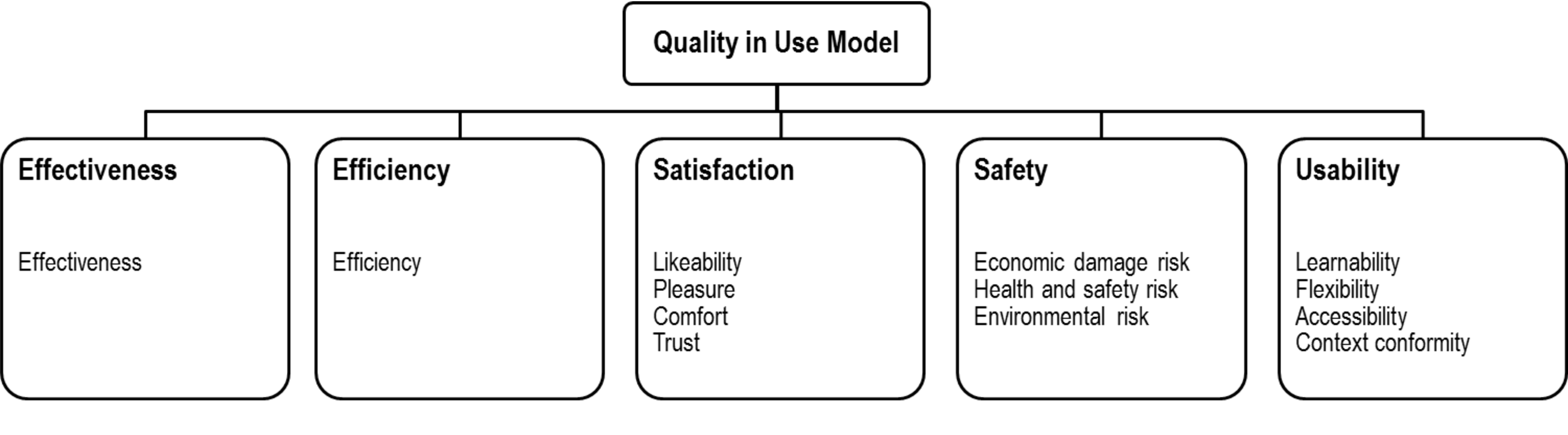}
	\caption{ISO 25010:2011 - Quality in use}
	\label{fig:QualityInUse}
\end{figure}

The quality in use model covers the outcome of the interaction when a product is used in a particular context of use providing a different perspective on quality. In our paper, we are interested in how quality in use is supported by MDE. The model is presented in figure \ref{fig:QualityInUse} and is also specified in terms of quality attributes and sub-attributes. Again, we will use the quality attributes to guide our discussion on how quality is supported in MDE.

\begin{itemize}
\item \textit{Effectiveness} is the capability to enable users to achieve specified goals with accuracy and completeness in a specified context of use.
\item \textit{Efficiency} is the capability to enable users to achieve specified goals in a timely manner, in a specified context of use.
\item \textit{Satisfaction} is the capability to satisfy users in a specified context of use.
\item \textit{Safety} is the capability to achieve acceptable levels of risk of harm to people, software, equipment or the environnent in a specified context of use.
\item \textit{Usability} is the capability to be easily used in a specified context of use.
\end{itemize}

\section{Results}
\label{sec:Results}

\subsection{Search results}

\begin{figure*}[ht!]
\centering
\subfigure[Papers by year]{%
{\includegraphics[scale=0.50]{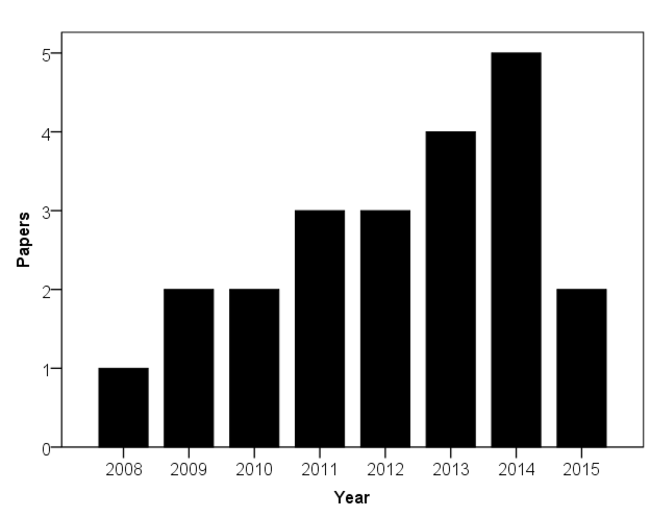}}%
\label{fig:PapersByYear}%
}\qquad
\subfigure[Included primary studies by review]{%
{\includegraphics[scale=0.23]{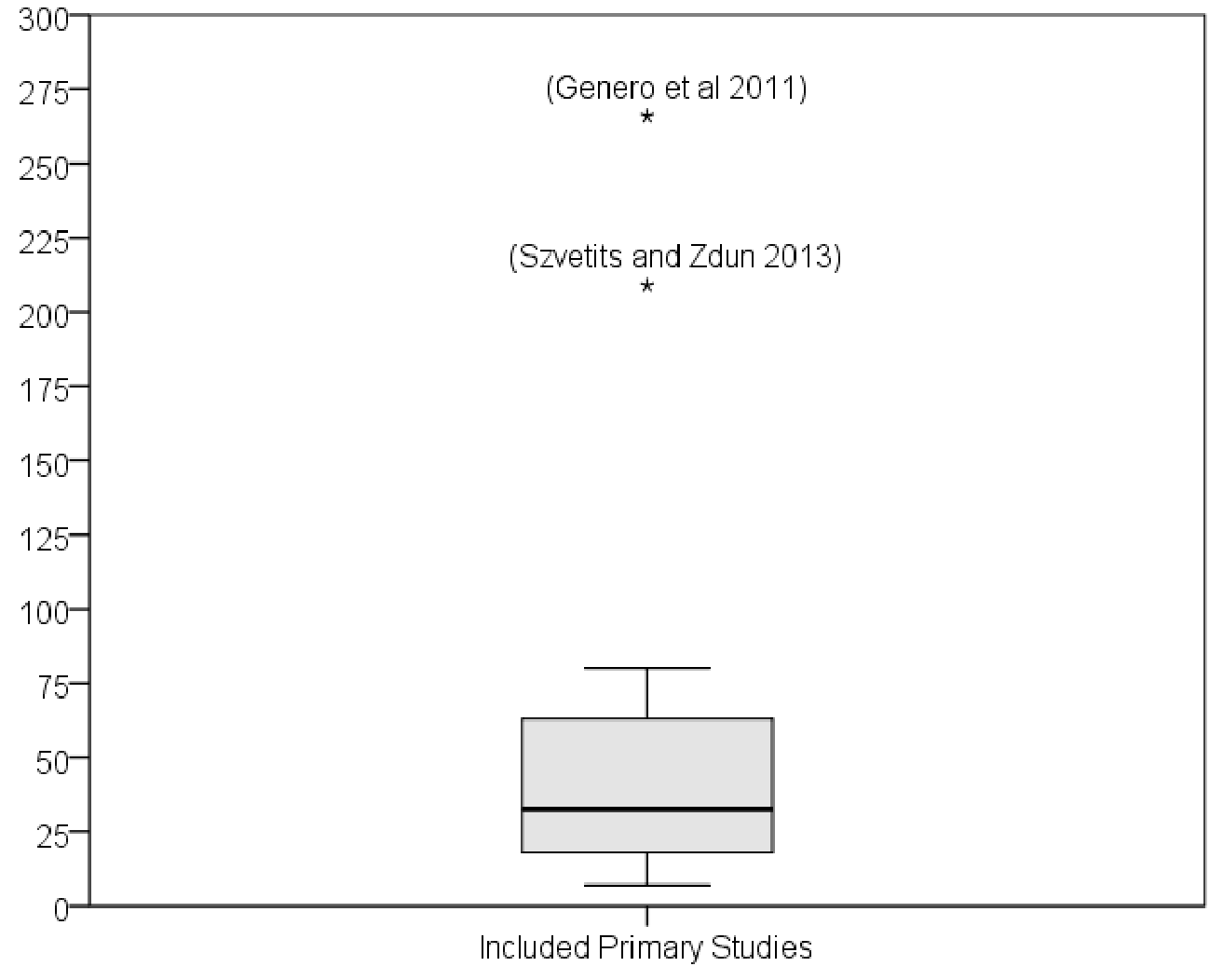}}%
\label{fig:IncludedPrimaryStudies}%
}
\subfigure[Citations by year]{
{\includegraphics[trim = 0.0cm 17.0cm 2.0cm 0.0cm, clip, width=0.85\columnwidth]{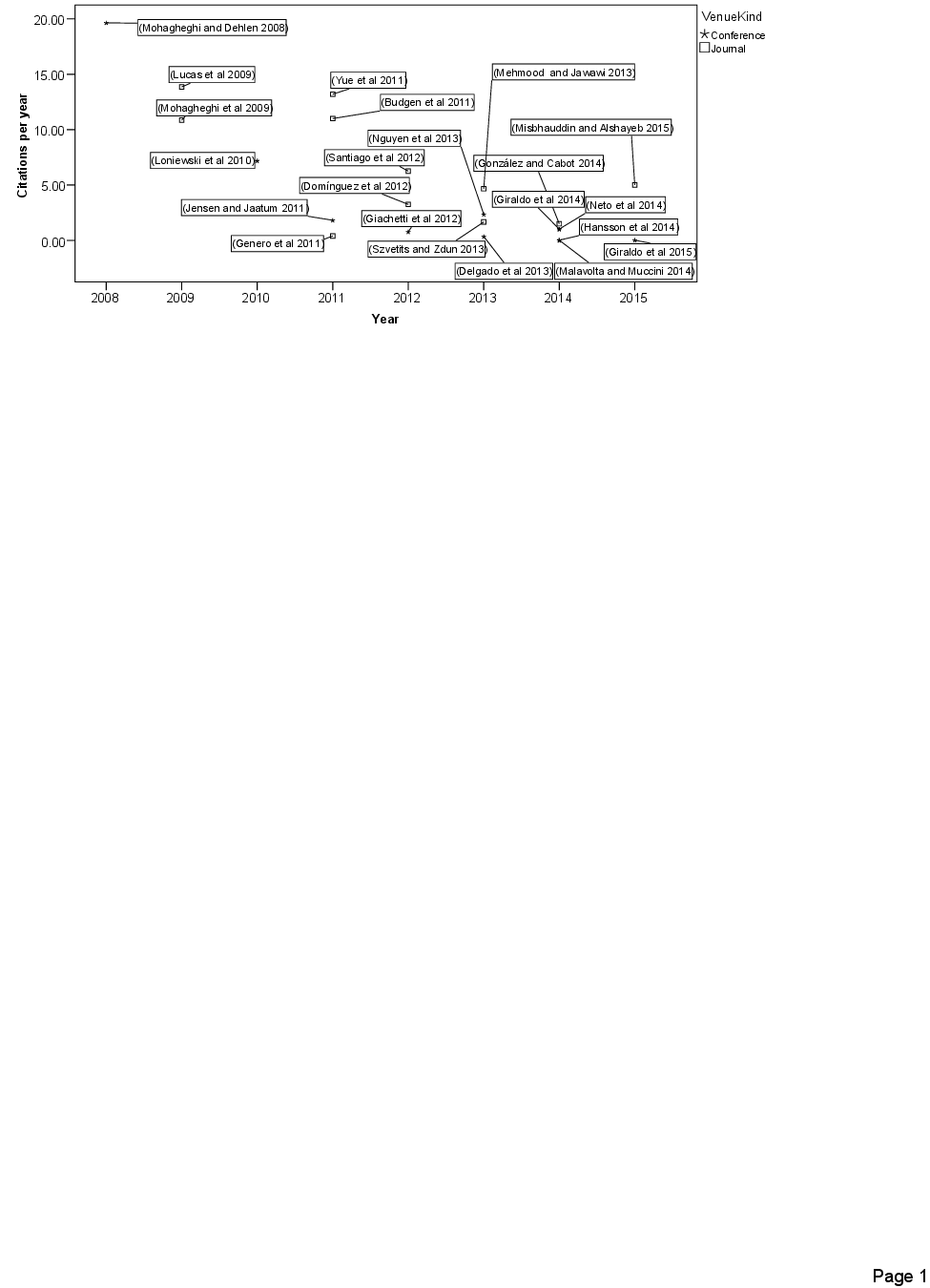}}
	\label{fig:CitationsPerYear}
}
\subfigure[Papers by team]{%
{\includegraphics[scale=0.25]{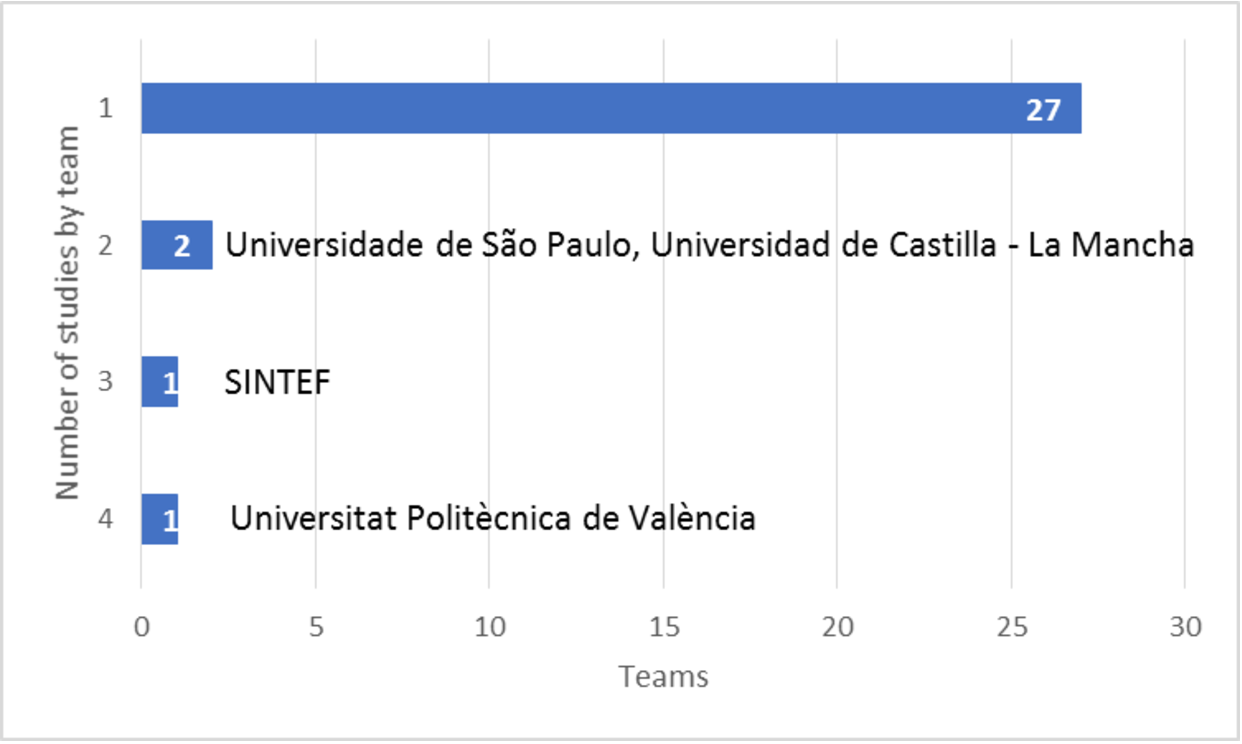}}%
\label{fig:PapersByTeam}%
}\qquad
\subfigure[Papers by author]{%
{\includegraphics[scale=0.25]{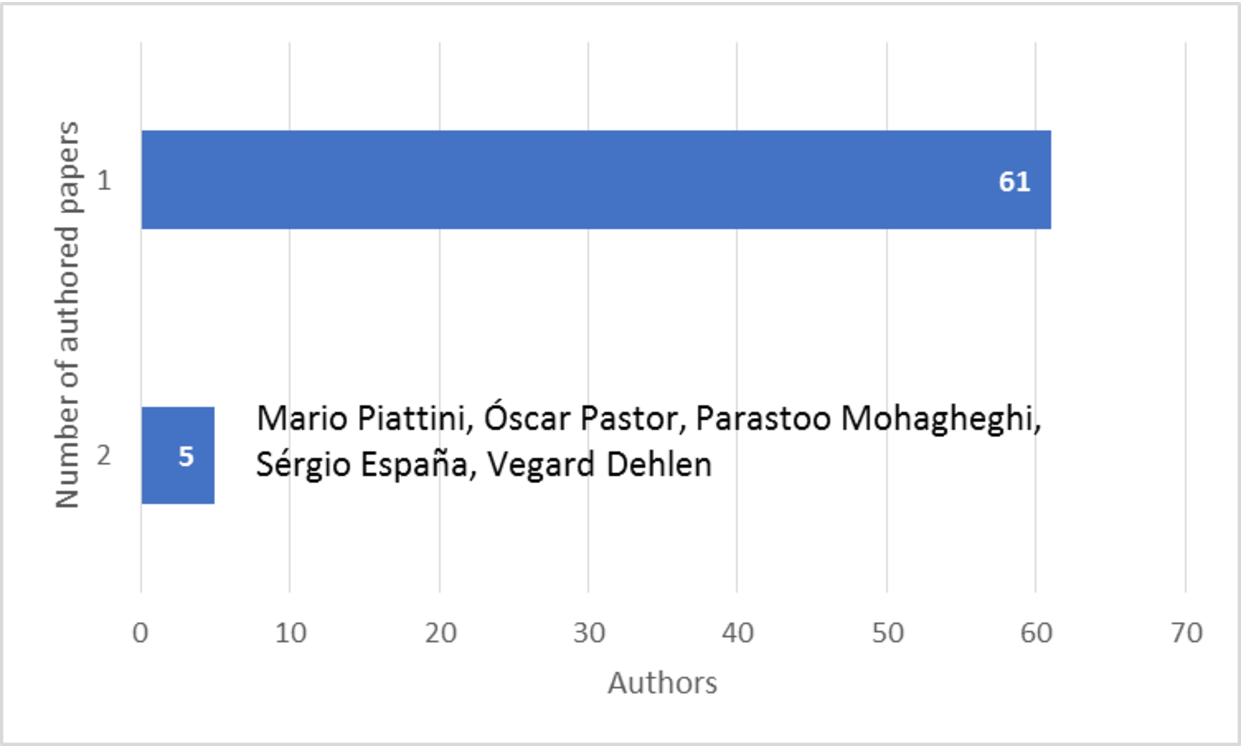}}
\label{fig:PapersByAuthor}%
}
\subfigure[Papers by country]{%
{\includegraphics[scale=0.3]{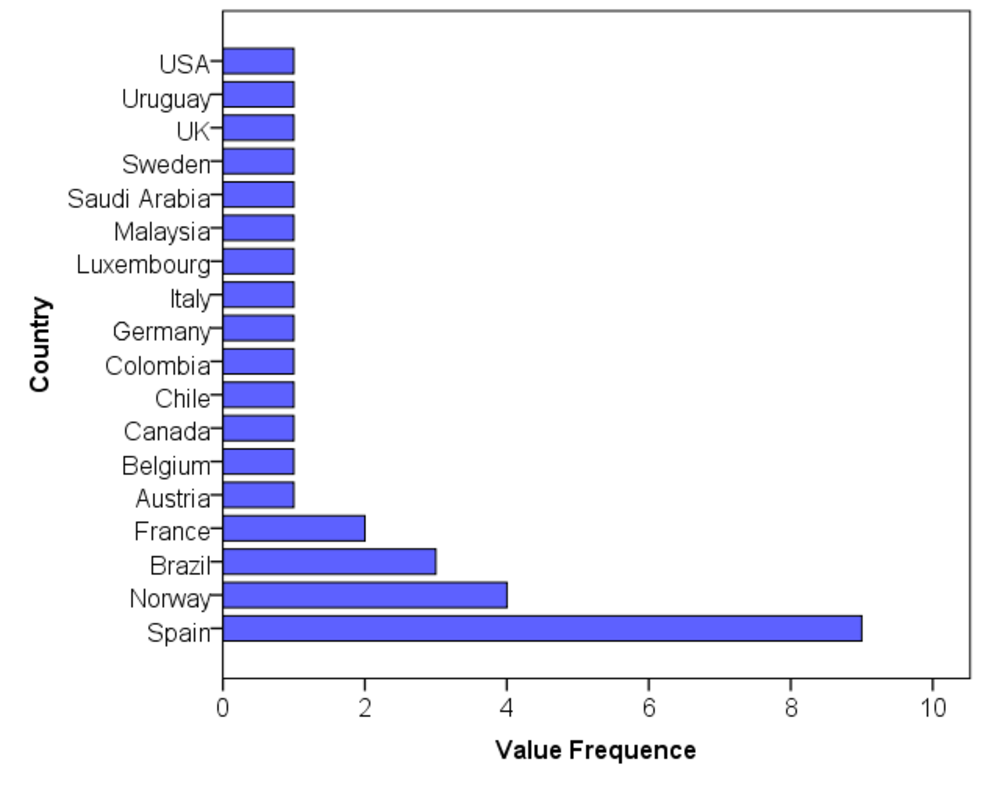}}%
\label{fig:PapersByCountry}%
}
\caption{Demographics informations on selected papers.}
\end{figure*}

We selected 22 papers for inclusion in this tertiary study.  There is an overall increasing trend in the number of secondary studies published covering this topic, which denotes a growing awareness on the subject (figure \ref{fig:PapersByYear}). Note that the data was collected in mid 2015, which may explain why there is a relatively low number of publications in that year. 
The distribution of the number of included primary studies has a mean of 53, a standard deviation of 63.4 with a minimum of 7 and a maximum of 266 included primary studies, in a total of 1166 included primary studies (without considering repetitions among different citations lists). The boxplot diagram in figure \ref{fig:IncludedPrimaryStudies} highlights two extremes, representing two secondary studies with a very high number of included primary studies, when compared to the other secondary studies.

The impact of the selected publications can be assessed through the assessment of the number of citations of each paper. As of July 16, 2015, the selected papers were cited 585 times, according to Google Scholar, with a mean of around 26.6 citations per paper, and a standard deviation of 40.7. Older papers have in general more citations than more recent ones. Figure \ref{fig:CitationsPerYear} presents the number of citations per paper divided by the paper's age, in years, distributed by year of publication and encoded by publication type (journal vs. conference paper). We present the average, rather than the total number of citations per paper to mitigate the cumulative effect of the paper's age in its number of citations. Journal papers tend to have more impact than conference papers, in this sample. The selected paper's set \textit{h-index} is 9. The \textit{h-index} is the largest number \textit{h} such that \textit{h} publications have at least \textit{h} citations \citep{Hirsch2005index}. The \textit{h-index} is often used to characterize the scientific output of a researcher. In this case, we use it to assess the impact of the body of publications included in this tertiary study.

We also characterize the community performing these studies.
The highest frequency of papers by institution is from Universitat Polit\`ecnica de Val\`encia - Spain (4), SINTEF - Norway (3), Universidade de S\~ao Paulo - Brazil (2) and Universidad de Castilla - La Mancha - Spain (2). All the remaining institutions contribute with 1 paper (figure \ref{fig:PapersByTeam}). 5 out of 66 authors have co-authored 2 of the selected papers each. The remaining 61 authors have only participated in one of the selected secondary studies (figure \ref{fig:PapersByAuthor}). 
The topic is widely distributed, with researchers from 4 continents, 18 countries and 32 institutions, and Spain as the most active country conducting secondary studies that cover quality in MDE (figure \ref{fig:PapersByCountry}). Overall, there is a clear predominance of European teams, followed by teams from South America. Concerning international collaboration among organizations from different countries, one paper \citep{Genero2011JDM} has authors from organizations from 3 different countries, while 6 papers have authors from organizations from 2 different countries. The remaining papers are authored by members of organizations from a single country. Again, Spain is the country involved in more international collaborations, participating in 4 out of 7 international teams.

Table \ref{tab:AudienceAimObject} summarizes the information on the intended \textit{audience}, \textit{aim} of the consolidated information, and \textit{object of study} of the selected secondary studies. Concerning the papers' audiences, here is a clear predominance of papers targeted to \textit{researchers} in the selected secondary studies. All the selected papers are targeted to an academic audience. 5 of these papers are also targeted at \textit{practitioners}. We identified 3 types of \textit{aim} of the information consolidated in the selected papers. \textit{Decision support} papers are the least common, with only 3 occurrences in our selection. This contrasts with 17 papers consolidating information relevant for \textit{knowledge support}. All the selected secondary studies have \textit{scoping} value for the respective areas. In other words, they provide an overview of the research area, collecting research evidence on the specific topic they address. Finally, we identified 3 main objects of study considered in these papers. 20 of these papers consolidate quality information about \textit{models}. 14 papers cover model transformations. 10 papers discuss the quality of MDE-related software processes.

\begin{table}[!ht]
\centering
\caption{Audience, aim and object of study of included studies}
\label{tab:AudienceAimObject}
\resizebox{\textwidth}{!}{%
\begin{tabular}{@{}lllllllll@{}}
\toprule
 & \multicolumn{2}{l}{Audience} & \multicolumn{3}{l}{Review aim} & \multicolumn{3}{l}{Object of study} \\ 
Paper Id & Researchers & Practitioners & Decision & Knowledge & Scoping & Models & Transformations & Processes \\
\midrule
\citep{Budgen2011SPE} & \checkmark & \checkmark &   & \checkmark & \checkmark & \checkmark &   &   \\
\citep{Delgado2013CCIS} & \checkmark &   &   & \checkmark & \checkmark & \checkmark & \checkmark & \checkmark \\
\citep{Dominguez2012IST} & \checkmark & \checkmark & \checkmark & \checkmark & \checkmark & \checkmark & \checkmark &   \\
\citep{Genero2011JDM} & \checkmark &   &   &   & \checkmark & \checkmark &   &   \\
\citep{Giachetti2012RCIS} & \checkmark &   &   & \checkmark & \checkmark & \checkmark & \checkmark & \checkmark \\
\citep{Giraldo2014RCIS} & \checkmark &   &   & \checkmark & \checkmark & \checkmark &   & \checkmark \\
\citep{Giraldo2015RCIS} & \checkmark & \checkmark &   &   & \checkmark & \checkmark & \checkmark & \checkmark \\
\citep{Gonzalez2014IST} & \checkmark &   &   & \checkmark & \checkmark & \checkmark &   &   \\
\citep{Hansson2014XM} & \checkmark & \checkmark & \checkmark & \checkmark & \checkmark &   &   & \checkmark \\
\citep{Jensen2011ARES} & \checkmark &   &   & \checkmark & \checkmark & \checkmark &   &   \\
\citep{Loniewski2010MODELS} & \checkmark &   &   & \checkmark & \checkmark & \checkmark & \checkmark &   \\
\citep{Lucas2009IST} & \checkmark &   &   & \checkmark & \checkmark & \checkmark & \checkmark &   \\
\citep{Malavolta2014ECSEAA} & \checkmark &   &   & \checkmark & \checkmark & \checkmark &   &   \\
\citep{Mehmood2013IST} & \checkmark &   &   &   & \checkmark & \checkmark & \checkmark & \checkmark \\
\citep{Misbhauddin2015EMSE} & \checkmark &   &   &   & \checkmark & \checkmark & \checkmark &   \\
\citep{Mohagheghi2008ECMDA-FA} & \checkmark & \checkmark &   & \checkmark & \checkmark & \checkmark & \checkmark & \checkmark \\
\citep{Mohagheghi2009IST} & \checkmark &   &   & \checkmark & \checkmark & \checkmark &   &   \\
\citep{Neto2014ECSAW} & \checkmark &   &   & \checkmark & \checkmark & \checkmark & \checkmark & \checkmark \\
\citep{Nguyen2013APSEC} & \checkmark &   &   &   & \checkmark & \checkmark & \checkmark & \checkmark \\
\citep{Santiago2012IST} & \checkmark &   &   & \checkmark & \checkmark & \checkmark & \checkmark & \checkmark \\
\citep{Szvetits2013SoSyM} & \checkmark & \checkmark &   & \checkmark & \checkmark & \checkmark & \checkmark &   \\
\citep{Yue2011JRE} & \checkmark &   & \checkmark & \checkmark & \checkmark &   & \checkmark &   \\ 
\midrule
Total & 22 & 6 & 3 & 17 & 22 & 20 & 14 & 10\\
\bottomrule
\end{tabular}
}
\end{table}

\subsection{Quality evaluation of secondary studies}
\label{subsec:ResultsQualityEval}

We applied the DARE quality criteria introduced in section \ref{subsec:MethodQualityAssessment} to all the selected papers. For the purposes of these classifications, we count criteria which are, at least, partially fulfilled. Finally, we also present the sum of all these quality criteria scores. The DARE \textbf{Q}uality \textbf{C}riteria are as follows:

\begin{itemize}
\item \textbf{QC1: Were inclusion and exclusion criteria reported?} To answer this question, we use a three-level scale. If the inclusion and exclusion criteria are explicitly reported, we encode this as 1. If these criteria are implicit, but can be safely inferred, we encode this as partial, corresponding to a score of 0.5. Finally, if the inclusion and exclusion criteria are absent from the paper, we grade this with 0.

\item \textbf{QC2: Was the search adequate?} To answer this question, we use a three-level scale. If the study reports on using 4 or more data repositories in its automated search process and an additional search mechanism, such as snowball search, we mark this as 1. If the study reports on using 3 or 4 data data repositories with no extra search strategies, we mark it as 0.5. Finally, if previous two conditions were not fulfilled we mark this as 0.

\item \textbf{QC3: Were the included studies synthesized?} We use a two-level scale for this question, with 1 standing for ``Yes'', and 0 for ``No''.

\item \textbf{QC4: Was the validity of the included studies synthesized?} We use a three level scale for this criterion. We assign 1 to studies where the authors explicitly defined quality or validity criteria and extracted them from the primary study. We assign 0.5 to studies where the research question involves quality or validity issues that are addressed by the study. We assign 0 if there is no explicit quality or validity assessment of the papers included in the secondary study.

\item \textbf{QC5: Are sufficient details about the individual studies presented?} We use a three level scale for this criterion. We assign 1 to papers when we can trace relevant information presented in a secondary study back to the primary study where it was collected from. We assign 0.5 when only summary information is presented about the individual papers. For example, papers may be grouped into specific categories, but it may not be explicit which papers belong to which specific category. Finally, we assign 0 when the results of individual studies are not specified.
\end{itemize}

With only two exceptions \citep{Genero2011JDM, Malavolta2014ECSEAA}, all the other studies meet at least the simplified DARE criteria, which are better suited to evaluate SLRs than SMSs. We opted to keep these two papers in our selected secondary studies as both are, in our opinion\footnote{Although \citep{Genero2011JDM} was published as a SLR, its 5 research questions are of a mapping nature. Under these circumstances, the quality classifications, which can be seen as \textit{adequacy for inclusion} classifications can be less strict.}, best characterized as SMSs, which partially explains their relatively lower score. Evaluation criteria such as the search adequacy and the quality evaluation of the included primary studies are less critical in SMSs. Overall, the quality of included secondary studies is very good.

\begin{table}[ht]
\centering
\caption{Included secondary studies quality assessment}
\label{my-label}
\resizebox{\textwidth}{!}{%
\begin{tabular}{@{}lllllllll@{}}
\toprule
Secondary study& QC1 & QC2 & QC3 & QC4 & QC5 & Total & Simple DARE & Full DARE\\
\midrule
\citep{Budgen2011SPE}& 1 & 1 & 1 & 0 & 1 & 4 & 1 & 1\\
\citep{Delgado2013CCIS}& 0.5 & 0.5 & 1 & 0 & 1 & 3 & 1 & 1\\
\citep{Dominguez2012IST}& 1 & 1 & 1 & 0 & 1 & 4 & 1 & 1\\
\citep{Genero2011JDM}& 1 & 0.5 & 0 & 0 & 0 & 1.5 & 0 & 0\\
\citep{Giachetti2012RCIS}& 1 & 1 & 1 & 1 & 1 & 5 & 1 & 1\\
\citep{Giraldo2014RCIS}& 1 & 1 & 1 & 1 & 1 & 5 & 1 & 1\\
\citep{Giraldo2015RCIS}& 1 & 0.5 & 1 & 0 & 1 & 3.5 & 1 & 1\\
\citep{Gonzalez2014IST}& 1 & 1 & 1  & 0.5 & 1 & 4.5 & 1 & 1\\
\citep{Hansson2014XM}& 0.5 & 0.5 & 1 & 1 & 1 & 4 & 1 & 1\\
\citep{Jensen2011ARES}& 1 & 0.5 & 1 & 0.5 & 1 & 4 & 1 & 1\\
\citep{Loniewski2010MODELS}& 1 & 1 & 0 & 1 & 1 & 4 & 1 & 1\\
\citep{Lucas2009IST}& 1 & 1 & 1 & 0 & 1 & 4 & 1 & 1\\
\citep{Malavolta2014ECSEAA}& 1 & 0 & 1 & 0 & 1 & 3 & 0 & 0\\
\citep{Mehmood2013IST}& 1 & 1 & 1 & 0 & 1 & 4 & 1 & 1\\
\citep{Misbhauddin2015EMSE}& 1 & 0.5 & 1 & 1 & 1 & 4.5 & 1 & 1\\
\citep{Mohagheghi2008ECMDA-FA}& 0.5 & 0.5 & 1 & 0 & 1 & 3 & 1 & 1\\
\citep{Mohagheghi2009IST}& 1 & 1 & 1 & 0 & 1 & 4 & 1 & 1\\
\citep{Neto2014ECSAW}& 1 & 0.5 & 1 & 1 & 1 & 4.5 & 1 & 1\\
\citep{Nguyen2013APSEC}& 1 & 1 & 1 & 0 & 0 & 3 & 1 & 0\\
\citep{Santiago2012IST}& 1 & 1 & 1 & 1 & 1 & 5 & 1 & 1\\
\citep{Szvetits2013SoSyM}& 1 & 0.5 & 1 & 0 & 1 & 3.5 & 1 & 1\\
\citep{Yue2011JRE}& 1 & 1 & 1 & 1 & 1 & 5 & 1 & 1\\
\bottomrule
\end{tabular}
}
\end{table}

All the selected secondary studies were built using at least one guideline, or a combination of guidelines. \citep{Kitchenham2007SR} is the most frequently used set of guidelines, similar with what we find in recent secondary and tertiary studies. A particularly noteworthy exception to the rule is the primary studies identification strategy followed in \citep{Giraldo2015RCIS}. Rather than essentially using a search-based approach to detecting primary studies to be potentially included, which may then be complemented with other strategies, as suggested in \citep{Kitchenham2007SR}, \citep{Giraldo2015RCIS} use the backward snowball approach, which was recently evaluated as an effective alternative for performing literature reviews in \citep{Wohlin2014EASE}.

\begin{table}[ht]
\centering
\caption{Guidelines used in the secondary study}
\label{tab:Guidelines}
\resizebox{\textwidth}{!}{%
\begin{tabular}{@{}llllllllll@{}}
\toprule
Paper Id & \begin{sideways}\tiny{\citep{Biolchini2005TR}}\end{sideways} & \begin{sideways}\tiny{\citep{Brereton2007JSS}}\end{sideways} & \begin{sideways}\tiny{\citep{Kitchenham2004ICSE}}\end{sideways} & \begin{sideways}\tiny{\citep{Kitchenham2007SR}}\end{sideways} & \begin{sideways}\tiny{\citep{Kitchenham2009IST}}\end{sideways} & \begin{sideways}\tiny{\citep{Mohagheghi2007EMSE}}\end{sideways} & \begin{sideways}\tiny{\citep{Petersen2008EASE}}\end{sideways} & \begin{sideways}\tiny{\citep{Petticrew2008Book}}\end{sideways} & \begin{sideways}\tiny{\citep{Wohlin2014EASE}}\end{sideways}\\ \midrule
\scriptsize{\citep{Budgen2011SPE}} &  &  &  & \checkmark &  &  &  &  &  \\
\scriptsize{\citep{Delgado2013CCIS}} &  &  & \checkmark & \checkmark &  &  &  &  &  \\
\scriptsize{\citep{Dominguez2012IST}} &  &  & \checkmark & \checkmark &  &  &  &  &  \\
\scriptsize{\citep{Genero2011JDM}} &  & \checkmark & \checkmark & \checkmark &  &  &  &  &  \\
\scriptsize{\citep{Giachetti2012RCIS}} &  &  &  & \checkmark &  &  &  & \checkmark &  \\
\scriptsize{\citep{Giraldo2014RCIS}} &  &  &  & \checkmark &  &  &  &  &  \\
\scriptsize{\citep{Giraldo2015RCIS}} &  &  &  &  &  &  &  &  & \checkmark \\
\scriptsize{\citep{Gonzalez2014IST}} &  &  & \checkmark & \checkmark &  &  &  &  &  \\
\scriptsize{\citep{Hansson2014XM}} &  &  &  & \checkmark &  &  &  &  &  \\
\scriptsize{\citep{Jensen2011ARES}} &  &  & \checkmark &  &  &  &  &  &  \\
\scriptsize{\citep{Loniewski2010MODELS}} &  &  & \checkmark &  &  &  &  &  &  \\
\scriptsize{\citep{Lucas2009IST}} &  &  &  & \checkmark &  &  &  &  &  \\
\scriptsize{\citep{Malavolta2014ECSEAA}} &  &  &  & \checkmark &  &  &  &  &  \\
\scriptsize{\citep{Mehmood2013IST}} &  &  &  & \checkmark &  &  & \checkmark &  &  \\
\scriptsize{\citep{Misbhauddin2015EMSE}} &  &  &  & \checkmark &  &  &  &  &  \\
\scriptsize{\citep{Mohagheghi2008ECMDA-FA}} &  &  &  &  &  & \checkmark &  &  &  \\
\scriptsize{\citep{Mohagheghi2009IST}} &  &  &  & \checkmark &  &  &  &  &  \\
\scriptsize{\citep{Neto2014ECSAW}} &  &  &  & \checkmark &  &  &  &  &  \\
\scriptsize{\citep{Nguyen2013APSEC}} &  &  &  & \checkmark &  &  &  &  &  \\
\scriptsize{\citep{Santiago2012IST}} & \checkmark &  &  & \checkmark &  &  &  &  &  \\
\scriptsize{\citep{Szvetits2013SoSyM}} &  &  &  & \checkmark & \checkmark &  &  &  &  \\
\scriptsize{\citep{Yue2011JRE}} &  &  &  & \checkmark &  &  &  &  &  \\
\midrule
Totals & 1 & 1 & 6 & 18 & 1 & 1 & 1 & 1 & 1 \\ \bottomrule
\end{tabular}
\normalsize
}
\end{table}

All but \citep{Nguyen2013APSEC} make the list of included primary studies available, either directly on the paper, or through an online resource (e.g. a companion site for the paper, or a more detailed technical report). 9 secondary studies explicitly assess the quality of the primary studies they include. 8 of those secondary studies use an \textit{ad-hoc} assessment, with custom-made criteria tailored for the specific domain those reviews are covering \citep{Delgado2013CCIS, Dominguez2012IST, Giachetti2012RCIS, Giraldo2014RCIS, Hansson2014XM, Misbhauddin2015EMSE, Mohagheghi2008ECMDA-FA, Santiago2012IST}. \citep{Neto2014ECSAW} uses guidelines \citep{Kitchenham2007SR} for the quality assessment of the included primary studies.

\subsection{Findings}

\subsubsection{On the nature of the research questions}
\begin{figure}[ht]
    \centering
    \includegraphics[scale=0.4]{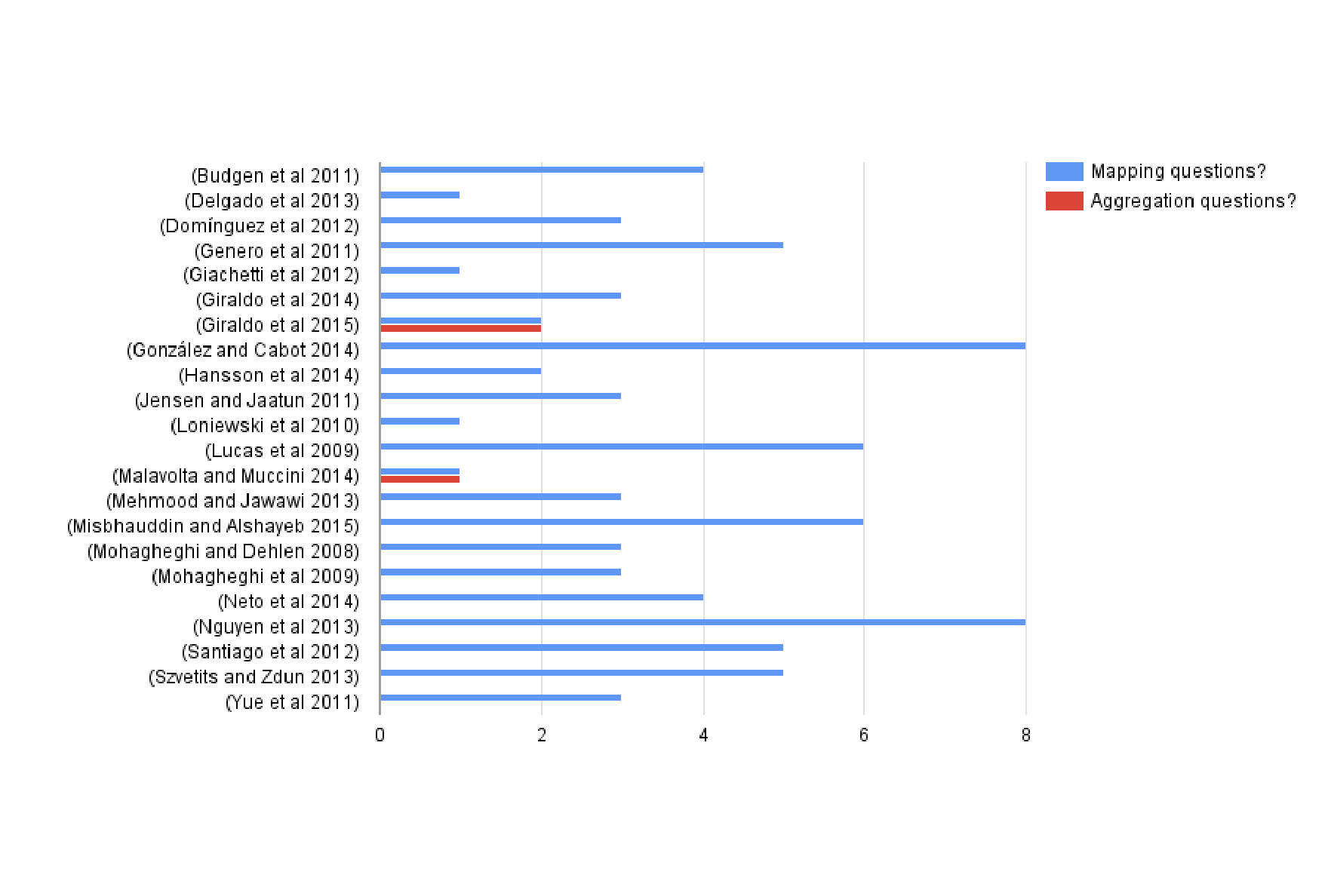}
    \caption{Research questions categorization}
    \label{fig:ResearchQuestionsType}
\end{figure}

Each secondary study has its own set of research questions. We identify two major kinds of research questions in these papers: those which are primarily focused in analysing trends in a particular quality in MDE toping area (mapping questions), and those aggregating data from different studies, typically to compare several alternatives, according to a specific set of quality attributes (aggregation questions).
80 of the research questions on the selected secondary studies have essentially a mapping nature (figure \ref{fig:ResearchQuestionsType}).
Only 3 research questions are of the form \textit{``How does A compare to B?''}, where \textit{A} and \textit{B} are MDE tools, techniques, processes, etc.

\subsubsection{On the coverage of quality attributes by the secondary studies}
Table \ref{tab:productQuality} summarizes the \textit{ISO 25010:2011 - Software product quality} attributes, and their coverage in the secondary studies included in this review. All the papers address at least one of the quality attributes. The most commonly addressed quality attribute is maintainability with 18 out of the 22 secondary studies. All the remaining quality attributes are addressed by a relatively low number of primary studies with 5 papers addressing reliability, efficiency and usability, 4 papers covering security, compatibility and portability, and 3 papers covering functional suitability.

\begin{table}[!ht]
\centering
\caption{Coverage of ISO 25010:2011 - Software product quality}
\label{tab:productQuality}
\resizebox{\textwidth}{!}{%
\begin{tabular}{@{}llllllllll@{}}
\toprule
Paper Id & \begin{tabular}[c]{@{}l@{}}Product\\ Quality\end{tabular} & \begin{tabular}[c]{@{}l@{}}Functional\\ Suitability\end{tabular} & Reliability & Efficiency & Usability & Security & Compatibility & Maintainability & Portability \\ \midrule
\citep{Budgen2011SPE} & \checkmark &  &  &  &  &  &  & \checkmark &  \\
\citep{Delgado2013CCIS} & \checkmark &  & \checkmark &  &  &  &  & \checkmark &  \\
\citep{Dominguez2012IST} & \checkmark &  &  & \checkmark & \checkmark &  &  & \checkmark &  \\
\citep{Genero2011JDM} & \checkmark & \checkmark & \checkmark &  &  &  &  & \checkmark &  \\
\citep{Giachetti2012RCIS} & \checkmark &  &  &  &  &  & \checkmark & \checkmark &  \\
\citep{Giraldo2014RCIS} & \checkmark &  &  &  &  &  &  & \checkmark &  \\
\citep{Giraldo2015RCIS} & \checkmark &  & \checkmark & \checkmark & \checkmark &  & \checkmark & \checkmark & \checkmark \\
\citep{Gonzalez2014IST} & \checkmark &  &  &  &  &  &  & \checkmark &  \\
\citep{Hansson2014XM} & \checkmark &  &  &  &  &  &  & \checkmark &  \\
\citep{Jensen2011ARES} & \checkmark &  &  &  &  & \checkmark &  &  &  \\
\citep{Loniewski2010MODELS} & \checkmark &  &  &  &  &  &  & \checkmark &  \\
\citep{Lucas2009IST} & \checkmark &  &  &  &  &  &  & \checkmark &  \\
\citep{Malavolta2014ECSEAA} & \checkmark &  &  & \checkmark &  &  &  & \checkmark &  \\
\citep{Mehmood2013IST} & \checkmark &  &  &  &  &  &  & \checkmark &  \\
\citep{Misbhauddin2015EMSE} & \checkmark &  &  &  &  &  &  & \checkmark &  \\
\citep{Mohagheghi2008ECMDA-FA} & \checkmark &  &  &  &  &  &  & \checkmark & \checkmark \\
\citep{Mohagheghi2009IST} & \checkmark & \checkmark &  &  & \checkmark &  &  & \checkmark &  \\
\citep{Neto2014ECSAW} & \checkmark &  & \checkmark &  & \checkmark & \checkmark & \checkmark &  & \checkmark \\
\citep{Nguyen2013APSEC} & \checkmark &  &  &  &  & \checkmark &  &  &  \\
\citep{Santiago2012IST} & \checkmark &  &  &  &  &  &  & \checkmark &  \\
\citep{Szvetits2013SoSyM} & \checkmark &  & \checkmark & \checkmark & \checkmark & \checkmark & \checkmark & \checkmark & \checkmark \\
\citep{Yue2011JRE} & \checkmark & \checkmark &  & \checkmark &  &  &  &  &  \\ \midrule
\textbf{Frequency} & 22 & 3 & 5 & 5 & 5 & 4 & 4 & 18 & 4\\
\bottomrule
\end{tabular}
}
\end{table}

Table \ref{tab:qualityInUse} summarizes the coverage of \textit{quality in use} attributes in the selected secondary studies. The quality in use is clearly a less explored view of quality, when compared to the product quality model. The model includes 5 attributes, of which the most explored is usability, with 7 occurrences. Effectiveness was explored in 2 studies. Finally, efficiency, satisfaction, and safety are the least explored quality in use attributes, being covered by 2 papers each. 9 papers do not cover quality in use at all. 
\begin{table}[ht]
\centering
\caption{Coverage of ISO 25010:2011 - Quality in Use}
\label{tab:qualityInUse}
\resizebox{\textwidth}{!}{%
\begin{tabular}{@{}lllllll@{}}
\toprule
Paper Id & Quality in use & Effectiveness & Efficiency & Satisfaction & Safety & Usability \\ \midrule
\citep{Budgen2011SPE} & \checkmark &  &  &  &  & \checkmark \\
\citep{Delgado2013CCIS} &  &  &  &  &  &  \\
\citep{Dominguez2012IST} & \checkmark &  &  &  &  & \checkmark \\
\citep{Genero2011JDM} &  &  &  &  &  &  \\
\citep{Giachetti2012RCIS} &  &  &  &  &  &  \\
\citep{Giraldo2014RCIS} & \checkmark &  &  & \checkmark &  & \checkmark \\
\citep{Giraldo2015RCIS} & \checkmark & \checkmark &  & \checkmark &  & \checkmark \\
\citep{Gonzalez2014IST} &  &  &  &  &  &  \\
\citep{Hansson2014XM} & \checkmark & \checkmark &  &  &  &  \\
\citep{Jensen2011ARES} &  &  &  &  &  &  \\
\citep{Loniewski2010MODELS} & \checkmark &  & \checkmark &  &  &  \\
\citep{Lucas2009IST} & \checkmark & \checkmark & \checkmark &  &  &  \\
\citep{Malavolta2014ECSEAA} & \checkmark & \checkmark &  &  &  &  \\
\citep{Mehmood2013IST} &  &  &  &  &  &  \\
\citep{Misbhauddin2015EMSE} & \checkmark &  &  &  &  & \\
\citep{Mohagheghi2008ECMDA-FA} &  &  &  &  &  &  \\
\citep{Mohagheghi2009IST} & \checkmark &  &  &  &  & \checkmark \\
\citep{Neto2014ECSAW} & \checkmark &  &  &  & \checkmark & \checkmark \\
\citep{Nguyen2013APSEC} &  &  &  &  &  &  \\
\citep{Santiago2012IST} &  &  &  &  &  &  \\
\citep{Szvetits2013SoSyM} & \checkmark &  &  &  & \checkmark & \checkmark \\
\citep{Yue2011JRE} & \checkmark &  &  &  &  &  \\ 
\midrule
\textbf{Frequency} & 13 & 4 & 2 & 2 & 2 & 7\\
\bottomrule
\end{tabular}
}
\end{table}

\subsection{Summary of secondary studies contributions}
\label{subsec:SecondaryStudiesContributions}
In this section, we briefly outline the main contributions of each of the included secondary studies. In some situations, where there is a strong connection between more than one SLR (or SMS), we merge their discussion, for easier comparison. For example, \citep{Budgen2011SPE, Genero2011JDM} are both about UML models; \citep{Giraldo2014RCIS, Giraldo2015RCIS} are closely related in that they explore the concept of quality, first in MDE in general, and then in MDE languages in particular.

\citep{Budgen2011SPE} presents a SLR on how the UML has been empirically evaluated. The authors concluded that \textit{comprehension} was the most widely studied category (with impacts on the maintainability and usability of the language), and that \textit{adoption} was an emerging topic which would interest practitioners. More generally, all topics could benefit from further research, particularly because there was a strong preponderance of lab evaluations, frequently performed with students, with relatively short durations, and a corresponding lack of field studies. Moreover, although UML has several different diagrams, there was a very strong predominance of studies covering class diagrams. Conversely, the remaining diagrams are insufficiently studied. An important conclusion of the SLR was that there was a clear need for more works to question its fitness for the intended purpose, rather than taking it for granted. 
\citep{Genero2011JDM} is also dedicated to the quality of UML models. In this case, the authors focus on conceptual models. The overall conclusions are compatible with those of \citep{Budgen2011SPE}. In particular, the authors emphasize the need for much more empirical validation of claims concerning UML conceptual models, and a much stronger interaction with practitioners in that validation. Most of the identified studies focus on semantic quality, but very few cover semantic completeness.

The methodological or conceptual applications of service-oriented computing and development, as well as \textbf{M}odel-\textbf{D}riven \textbf{D}evelopment (MDD) and MDE, to business processes and business process management are explored in an SLR reported in \citep{Delgado2013CCIS}. The SLR discusses the integration issues among these paradigms contrasting the studies which integrate business processes with model driven approaches, with service oriented approaches, and a combination of model-driven and service-oriented approaches. The latter is clearly the most commonly explored integration approach. The review discusses impacts of this integration which suggest positive effects on reliability and maintainability. 

State based languages, namely UML state machines, finite state machines and Harel state charts, can be used as a source for generating code. \citep{Dominguez2012IST} presents an SLR where some of the reviewed proposals discuss the positive impact on quality attributes such as reusability and maintainability. Most of the works build upon software design patterns. As an identified shortcoming, it is often the case that several proposals do not fully support important elements of rich state machine specifications. Another common shortcoming is that several proposals fail to provide an implementation strategy considering qualitative aspects in software development.

The multitude of MDD approaches and languages creates important challenges in terms of interoperability. \citep{Giachetti2012RCIS} surveyed this research area and found that although this was a hot topic, it was common to find several different approaches tackling essentially the same issues, but not relating them, particularly when it came to interoperability approaches focusing on supporting MDD processes. The authors found challenges in reaching a consensus in terminology and concepts used, which is a crucial step towards the desired interoperability.

\citep{Giraldo2014RCIS} discuss how the concept of quality is tackled in MDE literature. They identify a plethora of quality in MDE definitions (and some misconceptions, as well) and their associated trends, resulting from the different perspectives each of the primary studies is supporting. In an extension to this work, a second review is dedicated to the proliferation of modeling languages, and the quality of languages and models built with those languages. \citep{Giraldo2015RCIS} explore this topic and propose a research agenda, in order to cover several existing gaps in the current state of the art in modeling languages quality evaluation. In particular, they point to the growing complexity involved in MDE (in this case of an accidental nature, due to insufficient tool support, on the one hand, and to the lack of adequately trained personnel, on the other), the problem of dealing with several different complementary languages which may be used at the same, or different abstraction levels, and the importance of considering model transformations, as well.

When models are used as a starting point for a (semi-)automatic generation of the implementation, verifying them becomes crucial, as problems in the models can lead to undesired effects on the generated systems. \citep{Gonzalez2014IST} report on the existing approaches to formal verification of static MDD models. The approaches to verification in MDD typically start with a formalization step and are then followed by using the verification mechanisms available to the specific formalism used. This formalization is often materialized in UML class diagrams (or some other similar approach), complemented with \textbf{O}bject \textbf{C}onstraint \textbf{L}anguage (OCL). The completeness of these approaches is strongly influenced by the level of support to OCL in the used tools. The most commonly verified properties are satisfiability and relationships among constraints (e.g. non-redundancy). The authors suggest the creation of a catalog of commonly agreed verification properties, and a set of benchmarks with which the approaches claiming to support their verification can be checked. This review highlights the need for a stronger efficiency and a better integration of the verification tools on the development tool chain, facilitating a more widespread adoption.

The SLR on model-driven agile development \citep{Hansson2014XM} concludes that this research field is still immature with respect to empirical evidence and industrial experience. The main aim of integrating agile practices into MDD were involving stakeholders, improving productivity and quality, and shortening the lead time. Due to immaturity the successfulness of the approach is still not yet quantified.

The main findings of an SLR on security in MDD \citep{Jensen2011ARES} are based mainly on experience reports involving primarily prototypes. These primary studies provide little evidence that automatically generated code is more secure or better than those obtained by other development approaches. On the other hand it was reported that security design models are understandable and expressive enough to model different access control policies. The study also reveals that security experts must be able to evaluate the quality of transformation rules, which is often beyond their capabilities.

Some of interesting findings of SLR on the use of requirements engineering techniques in MDD are that about of 60\% of primary studies included in \citep{Loniewski2010MODELS} already used models as a means to represent requirements, and about 40\% of them perform fully automatic transition from requirements specifications to analysis and design. Interestingly, almost 60\% of included primary studies have some automated support for traceability management. However, a complete solution to manage models in requirements phase is still lacking. Furthermore, empirical studies showing enhanced productivity, efficiency and effectiveness are still missing.

Models are frequently specified with different languages, each providing its specialized view on the system. UML, for example, is composed of several different sub-languages, and the different model views provided by each language must be consistent among them. \citep{Lucas2009IST} reviews the state of the art of UML model consistency checking. One of the important outcomes is that vertical inconsistency problems were (at least back in 2009) much less explored by research than horizontal ones. The authors of the review also proposed an approach to mitigate the identified gaps in the existing research, building on models transformations and rewriting logic.

\citep{Malavolta2014ECSEAA} reports a SMS on how MDE can be used in the context of on wireless sensor networks. The authors identify the main motivations for researchers to use MDE in this research field. The main conclusion was that MDE is used due to automatic support of code generation and documentation, as well as support for different analysis (e.g., performance, fault tolerance, power consumption, security). Both have impact on different quality issues. However, there is still lack of a satisfactory standard language for modeling wireless sensor networks.

The key findings of a SMS on aspect-oriented model-driven code generation by \citep{Mehmood2013IST} are: although the most significant advantages of automatic model-driven code generation are reduction of development time and improvement in maintainability, extensibility, and reliability these have yet to be proved for aspect-oriented model-driven modeling since mainly solution proposals exist. Validation and evaluation of these proposal are rare indicating that model verification is harder since designers need to know the details of advice transformations resulting in usability problems; aspect-oriented features are less appropriate for complex modeling situations that require weaving beyond simple model composition and transformation (e.g., with complex join-points); and problems with dynamic aspect weaving and unweaving.

The effect of model refactoring on model quality has been one of the research questions in SLR \citep{Misbhauddin2015EMSE}. It turns out that only 5 primary studies out of 63 addressed this topic and research is clearly inadequate. The only technique applied is to use quality design metrics before and after refactoring. However, established correlation between these metrics and quality attributes is still missing.

Ultimately, the adoption of MDE in industry should be guided by a clear understanding of the involved benefits and limitations. \citep{Mohagheghi2008ECMDA-FA} reviewed publications from 2000 to 2007 covering industrial experiences with MDE and found that, at the time, third-party tool maturity was not satisfactory for large-scale adoption. The authors did find some evidence on the benefits of MDE adoption in industry, where improvements in software quality were accompanied with reports of productivity gains and losses. However, these reports were mostly based on small-scale projects, so studies on larger projects were necessary to strengthen the evidence. Overall, the gathered evidence was insufficient for generalization of results, particularly for large-scale projects.

\citep{Mohagheghi2009IST} presents a SLR  discussing the meaning of model quality and how to improve it. In this work, six model quality goals were identified: correctness, completeness, consistency, comprehensibility, confinement and changeability. Some extra findings are discussed such as, for instance, to manage changeability and complexity of large and complex models, keep them consistent, and verify quality on the model level are challenges in model-driven engineering that are not yet properly covered.

\citep{Neto2014ECSAW} addresses the usage of MDE in the development of \textbf{S}ystems \textbf{o}f \textbf{S}ystems (SoS). The authors found that UML and OCL are the most frequently reported languages used in this context. Eclipse-based tools built with EMF/GMF are the most commonly reported as supporting the construction of SoS. Although there is a wide coverage of quality attributes in this context, interoperability is the most frequent quality concern, followed by reliability, safety and security.

In the SMS presented in \citep{Nguyen2013APSEC} we can find a taxonomy focused on specialized Model-Driven Engineering approaches for supporting the development of systems regarding security (also called Model-Driven Security, MDS). This study shows that authorization, especially access control, and confidentiality are the security concerns that are mostly addressed by MDS. Regarding the modeling approaches, the separation of concerns is regarded as prevalent, and \textbf{D}omain-\textbf{S}pecific \textbf{L}anguages (DSLs) are a popular approach for leveraging MDE for secure systems development (despite the fact that the great majority of those DSLs are implemented using UML profiles and stereotypes). Also, a small fraction of those describe the semantics properly, which means that they are not meant for formal analysis. Finally, a residual part of the selected studies present some sort of evaluation mostly validating via illustrations.

\citep{Santiago2012IST} focuses on traceability management in the context of MDE. They observe that the most addressed traceability operations are storage, \textbf{c}reate, \textbf{r}ead, \textbf{u}pdate and \textbf{d}elete (CRUD), and visualization. However, often CRUD operations are only partially supported, and the visualization mechanisms are typically ad-hoc and do not fit well the specific needs of visualizing traceability information, according to the author's analysis. Another identified shortcoming is the predominance of technology dependent traceability links generation from model transformations which hampers their application to other existing model transformation languages. The traceability metamodel is also proposal-specific, which complicates information interchange.

Models at runtime cope with the systems dynamic aspects. \citep{Szvetits2013SoSyM} reviews and classifies the current research approaches with respect to their objectives (e.g. adaptation, abstraction, consistency, conformance, error handling, monitoring, simulation, prediction, platform independence, and policy checking and enforcement), techniques (e.g. introspection, model conformance, model comparison,
model transformation, and model execution), architectures (e.g. monolithic, local dataflow, middleware, repository, or a combination of several of these), and kinds of models used in these approaches. A common trend observed in the included primary studies is a concern shift from low-level system interactions to model-based processing, which is closer to the problem domain.

Focused on transformation approaches between user requirements and analysis models, paper \citep{Yue2011JRE} is a SLR that unveils that there is still not practical, workable and automated solution in this topic. Those existing ones require significant user effort to document requirements, or are inefficiency, or are not able to (semi-) automatically generate a complete and consistent analysis model. The authors suggest to keep researching on transformations focusing on empirical studies, their quality characteristics such as usability, efficiency, scalability, extensibility, and interoperability.
\section{Discussion}
\label{sec:Discussion}
In this section we revisit the research questions, in light of the data gathered in this tertiary review (sub-section \ref{subsec:DiscussionResearchResults}).
This discussion follows with the proposal of a research roadmap, in section \ref{subsec:roadmap}.
Finally, we discuss the limitations of this study and our mitigation strategies for each of them in section \ref{subsec:Limitations}.

\subsection{Discussion of the Research Questions}
\label{subsec:DiscussionResearchResults}

\noindent \textit{RQ1: What is the currently available information concerning quality in MDE, systematically aggregated by means of a SLR, or a SMS?}

\noindent There is already a relatively large body of work synthesized which is somewhat related to quality in MDE. Our review included 22 secondary studies with a mean number of 53 included primary studies. Overall, the selected secondary studies included a total of 1166 primary studies (RQ1.1). The primary studies span from 1990 to 2014 (RQ1.2), and their list is available for all but one of the included secondary studies (RQ1.3). However, in most cases, the quality of the primary studies is not explicitly assessed (RQ1.4). When this assessment is performed, it is mostly done \textit{ad-hoc}, with custom-made criteria for this assessment. In some cases, the assessment of the primary studies is performed following suggestions in \citep{Kitchenham2007SR}. 

The available secondary studies are mostly targeted to researchers -- all the selected secondary studies have researchers as their target audience (RQ1.5). Nevertheless, 6 of the secondary studies are also aimed at practitioners. This large predominance of research-oriented papers may be interpreted as a hint for the level of maturity of the research domains covered in the reviews, which may be perceived as mature enough so that a consolidation effort is necessary, but perhaps not yet at a point where this consolidation is at the adequate level for discussing the collected evidence with practitioners. An alternative explanation could be that researchers may not necessarily regard these secondary studies as the best vehicle for conveying the consolidated findings to practitioners. Further research on this topic is certainly necessary. This shortcoming of lack of advice targeted for practitioners is not exactly surprising. In a tertiary study on SLRs and SMSs in software engineering a similar problem was identified \citep{Kitchenham2009IST}. The original purpose of the Evidence-Based Software Engineering movement was to provide consolidated evidence to practitioners, as well, so more research targeting practitioners is necessary.  Whatever the reason is, we note that although all the included secondary studies have a scoping goal, and most of them are also concerned with knowledge consolidation, only three of those studies explicitly address decision support as one of their aims. Again, this may be a symptom of an insufficient orientation for practitioners of the results in these secondary studies (RQ1.6). 

Different aspects of MDE (and how quality relates to them) are covered in these secondary studies (RQ1.7): 20 of the included secondary reviews cover models; 14 cover transformations; finally, 10 cover processes. Although quality models are not explicitly discussed in the selected secondary studies, we found that all those studies discuss at least one aspect of software product quality (RQ1.8). The most frequently addressed quality attribute is maintainability. This is in line with one of the ``selling points'' of MDE: the potential economic benefits it may bring in terms of costs savings in the software process. Each of the remaining product quality attributes listed in the ISO/IEC 25010:2011 standard is explored in a few secondary studies (from 3 to 5, depending on the specific quality attribute). We also explored the ISO/IEC 25010:2011 standard quality in use attributes, and found usability to be the most frequently covered (in 7 of the 22 secondary studies). This has mostly to do with the readability and understandability of the models. In other words, the added value of the MDE approach in terms of communication (e.g. by increasing the understandability of the software product through models representing it) is the most explored quality in use feature. Again, this has a potential impact on the economics of MDE, and in particular, in the maintainability of software developed with MDE.

\medskip
\noindent \textit{RQ2: What is the current status of consolidation of data collected from different literature reviews covering quality in MDE?}

As can be seen from section \ref{subsec:SecondaryStudiesContributions}, there is little overlap in the focus of the different included secondary studies. In other words, the intersection of the object of study and targeted quality attributes in the different primary studies is nearly empty. This makes performing meta-analysis focusing on a particular niche of this domain harder, at a tertiary level. At the secondary level (the one of the included studies) we note that the research questions are predominantly of a ``mapping nature''. Out of 83 research questions collected overall, only 3 were not essentially mapping questions. Mapping questions are, of course, extremely important, as they help locating relevant research about each of the topics addressed in the secondary studies. However, the relative absence of questions of the form \textit{``How does A compare to B?''}, where \textit{A} and \textit{B} are MDE tools, techniques, processes, etc, denotes a relatively low level of adoption of meta-analysis. This is not uncommon in Software Engineering, in general. As a community, we still struggle to aggregate independently collected evidence. Although the Evidence-Based Software Engineering approach is becoming more widespread, its adoption is still relatively low when compared to other domains. The same general problem is observable in this particular context of quality in MDE. This shortcoming of the current state of the art is, in our opinion, a \textit{``growing pain''} in a maturing research area and should be regarded as an opportunity for much needed further research.

\medskip
\noindent \textit{RQ3: Who are the key players in consolidating knowledge on quality in MDE through literature surveys?}

There is a wide dispersion of authors performing these literature reviews, with only 5 out of 66 authoring more than one secondary study, and even those only authored two studies. In other words, there does not seem to be a well defined set of key players contributing to the consolidation of knowledge in quality in MDE, following this particular approach. Obviously, this is not likely to be the case in each software quality niche, as explored by the corresponding primary studies. We found 22 eligible secondary studies (RQ3.1) addressing the topic of quality in MDE. This is a highly significant number of secondary studies. For comparison, a comprehensive tertiary study (and with a larger scope) on how research synthesis is performed in software engineering, included 49 secondary studies published from 2005 to 2010 \citep{Cruzes2011IST}. This is partly explained by the increasing adoption of secondary studies in software engineering, but also reflects a vibrant search area, with respect to the topic of our paper. These studies are mostly conducted by non-intersecting teams spread around 4 continents, 18 countries and 32 institutions (RQ3.2). Overall, and perhaps unsurprisingly, rather than identifying a community specialized in conducting these reviews for this particular domain, we found a more \textit{``opportunistic''} community, in the sense that these teams spotted gaps in the consolidated knowledge on quality in MDE, in the scope of their research work, and took the opportunity to fulfill them with much needed secondary studies. This hints to the increasingly widespread adoption of this kind of reviews, which has clearly outgrown, as intended, the specialized Evidence-Based Software Engineering community. 

In terms of impact (RQ3.3), the selected publications have an \textit{h-index} of 9. Considering some of the included publications are quite recent, there is clearly an audience for these secondary studies.

\subsection{A research roadmap for Quality in MDE}
\label{subsec:roadmap}
As stated in Introduction the main aim of this tertiary study is collecting the evidence concerning the impact of MDE in quality. In particular, we were interested in the quality of products build with MDE and the quality of process using MDE. In this respect 22 secondary studies, SLRs and SMSs, have been surveyed to get an overview of how MDE impacts quality and to provide a roadmap for future research of quality in MDE. Although a substantial body of primary and secondary study exists (22 secondary studies which examined over 1000 primary studies) none of the secondary studies explicitly addresses quality attributes for product quality and quality in use. There is a clear \textbf{lack of secondary studies collecting evidence of possible enhanced quality brought by MDE}. 

Industry needs a clear evidence on MDE impacts on quality supported by controlled experiments in small, medium, and large environments. Findings from such primary studies then need to be synthesized by secondary studies providing a clear evidence for industry. This current lack of orientation towards practical use is very visible from our study. \textbf{All 22 secondary studies used in our tertiary study have been intended to use for researchers, only a few of them addressed also practitioners. }

We urge prospective researchers to \textbf{explicitly address quality attributes} in their secondary studies. Maintainability is among those software product quality attributes, which were most often implicitly addressed. Much less attention has been paid to other software product quality attributes, such as functional suitability, compatibility, performance efficiency, security, compatibility, and portability. Although some secondary studies provide a weak evidence on positive effects of MDE on maintainability, reusability and reliability, we need studies which will address these attributes explicitly. 

On the other hand, \textbf{quality in use attributes were less investigated than software product quality attributes} (all 22 secondary studies implicitly address at least one software product quality attributes, whilst only 13 secondary studies implicitly addressed quality in use attributes). There is a clear lack of primary studies (controlled experiments) and secondary studies, which will explicitly address and synthesize knowledge about quality in use attributes, such as: effectiveness, efficiency, satisfaction, safety, and usability. Industry needs evidence supported by secondary studies which will explicitly cover the impact of MDE on quality in use attributes. 

Last, but not least, our tertiary study clearly shows that \textbf{most of secondary studies consolidate quality information about models, much less about model transformations and software processes}. These gaps should be addressed in the near future. We encourage MDE researchers to start addressing the identified gaps to support practitioners with evidence on MDE impacts on quality.

\subsection{Limitations of this study}
\label{subsec:Limitations}
A first potential limitation of this study is the coverage of our search and selection process. It is possible that a relevant secondary study may have been missed, either for not being captured at all with our search strings, in the chosen repositories, or for being mis-classified as irrelevant, when in fact it was relevant. The fact that our initial set of identified studies was all captured by our search strings gives us some confidence on the coverage of the search strings, although we did identify 4 extra studies (1 of which was eventually discarded), not captured through the search strings, using snowball search. As for the potential for missing a paper within the set, it is always possible that an unclear, or mis-interpreted, title and abstract could lead to a wrong exclusion. To mitigate this threat, we were very conservative in our exclusion decisions. When in doubt, we kept the papers for further analysis by a different author. If the doubt persisted, the paper passed on to the next phase. 

A second threat, common to this kind of studies, is that there is always some potential for mis-interpreting the secondary study. Again, we tried to be conservative, used a common data collection tool (implemented as a shared spreadsheet), and when in doubt marked the specific data element for discussion with the remaining authors, so that a consensus interpretation could be reached.

Finally, the authors of the included secondary studies may have also had problems both in the selection and the interpretation of the primary studies. If wrong conclusions were reached in an included secondary study, these could propagate to our tertiary study. To gauge this threat, we used the DARE Quality Criteria to assess the quality of the included secondary studies. The included secondary studies were of high quality, giving us some confidence that this potential propagation of wrong conclusions did not occur. Also, we did not perform meta-analysis on the data collected from the secondary studies, which would have the potential effect of amplifying errors from the secondary studies included in this tertiary study. 
\section{Conclusions and further work}
\label{sec:Conclusions}
This paper contributes with a tertiary study on the quality in MDE. We identified 22 secondary studies (SLRs and SMSs). In these secondary studies, we looked for discussions on how MDE impacts different quality attributes, taken from the ISO standards quality attributes for product quality and for quality in use. The product quality was chosen so that we could analyse existing work in a wide range of internal and external product quality attributes. The quality in use was chosen for this tertiary study as we were also interested in learning about the current body of knowledge of the impact of MDE in the quality in use of products built with it.
We now revisit the main contributions of this paper.

The first contribution concerns mapping the most representative secondary studies that cover quality in MDE, their origin (to identify key players in this research area), and the quality attributes addressed by each of the secondary studies. The 22 studies mostly come from independent teams, each concerned with its own niche. The coverage of quality attributes is wide, but with a predominance of product quality attributes over quality in use attributes. This observation seems to be inline with the common claims of MDE having a positive impact in terms of costs reduction in software development, and not so much in the quality of the product as perceived by its end user, who typically does not even need to know how the product was built. Maintainability is, by far, the most often addressed quality attribute. In general the contributions are more often targeted to researchers than to practitioners, which are only clearly targeted by 6 of the reviews.

A second contribution of this paper is an annotated overview of the existing aggregated information on quality in MDE. This is intended to serve as a starting point for interested readers to explore particular niches, by directing them to a particular literature review.

The third contribution of this paper is an analysis on the level of consolidation of the aggregated information on quality in MDE. The large predominance of questions more aimed at mapping the current body of knowledge than to compare among two, or more, alternatives, is a symptom of the relative novelty of the area (when compared to more ``traditional'' sciences), which is reaching a maturity level where the secondary studies can progressively focus more on answering more specific research questions comparing among different alternatives. The wide dispersion of sub-topics addressed in the different reviews makes it hard to perform meta-analysis on the collected data, hampering some of the potential benefits of SLRs. 

Last, but not the least, we propose a research roadmap for Quality in MDE, based on the main findings of our tertiary study. The identified shortcomings of the current state of the art in this domain offer important opportunities for conducting relevant consolidation efforts and, consequently, pushing MDE envelope to practitioners who still may have doubts on the potential benefits for their specific contexts. 

\begin{acknowledgements}

The authors would like to thank FCT/MEC NOVA LINCS PEst UID/ CEC/04516/ 2013 for the financial support to this work.
\end{acknowledgements}

\bibliographystyle{spbasic}      
\bibliography{references}


\end{document}